\documentclass[aps,prd,
showpacs,amssymb,
amsfonts,epsf,preprintnumbers,nofootinbib,superscriptaddress]{revtex4}

\usepackage[dvips]{graphicx}
\usepackage{bm,latexsym,amsmath,amssymb,amsfonts,color
}

\newcommand{\nn}{\nonumber}

\newcommand{\ta}{\left(\frac{3H\tau}2\right)}

\begin{document}

\title{Evolution of cosmological event horizons in anisotropic universes }

\author{Hyeong-Chan Kim}
\email{hckim@cjnu.ac.kr}
\affiliation{School of Liberal Arts and Sciences, Chungju National University,
Chungju, 380-702, Korea}

\begin{abstract}
We study the evolution of cosmological event horizons in anisotropic Kasner universes in the presence of a positive cosmological constant by analyzing null geodesics.
At later times, the asymptotic form of cosmological horizons is the same spherical surface as the de Sitter horizon.
At the early times, however, it has non-spherical shape with its eccentricity decreases with time.
The horizon area increases with time respecting the second law of thermodynamics.
The initial shape of the cosmological horizon takes the form of a needle or pancake surface depending on the nature of the background spacetimes.
We also discuss that the presence of the holographic dark energy will modify significantly the initial evolution of the anisotropic universes.
\end{abstract}

\maketitle

Following the Copernican principle~\cite{Peacock}, the physical properties of the universe are spatially homogeneous viewed on a sufficiently large scale.
This is consistent with nowadays experiments up to of order $10^{-5}$.
Based on the principle, the Robertson-Walker spacetime with inflation at early stage, on which small perturbations evolve to give cosmological structures, is widely used to describe the present universe.
However, there is no reason to believe that our universe was initially homogenous and isotropic.
Generically, it could be highly inhomogeneous and anisotropic.
The cosmic no-hair theorem~\cite{CNHT} shows that in the presence of inflation most of the initially anisotropic cosmological spacetimes become isotropic rapidly if the matter fields residing in the universe satisfy the dominant energy condition.
The theorem also implies that it is inevitable to have a highly anisotropic pre-inflationary universe if any small anisotropy remains after inflation.
Since the anisotropy will be exponentially enhanced if we observe the universe backward in time and we cannot make that process stop by adding ordinary matters or radiations.
The vestige of the pre-inflationary anisotropies can be seen through the large scale anomalies in the WMAP power spectra if the $e$-folding of inflation is not larger than $64$ for GUT scale inflation.~\cite{kM1,kM2}.
In fact, there is possibility that the effect of pre-inflationary anisotropies has been already detected in the Cosmic Microwave Background (CMB) temperature anisotropies, as an apparent alignment of the CMB multipoles on very large scales, so-called ``Axis of Evil"~\cite{anomaly1,anomaly2,copi,cruz,eriksen,Picon} and giant rings~\cite{Kovetz}.

The standard cosmological model is the $\Lambda$-CDM model where most part of the universe is composed of the cosmological constant and cold dark matters.
The spacetime has a future event horizon located at a distance
$R_h =a(\tau)\int_\tau^\infty \frac{d\tau'}{a(\tau')},$ 
where $a(\tau)$ is the scale factor of the Robertson-Walker metric and $\tau$ is the comoving time.
In the presence of asymptotically accelerating expansion of the universe, any observers will be surrounded by the future event horizon, which restricts the observers access to the information of the universe.
This lack of information is represented by a kind of entropy proportional to the area of the cosmological horizon divided by the Planck area similar to the black hole entropy.
The generalized second law of thermodynamics for the Robertson-Walker spacetime was proved by Daives~\cite{Davies} and Pollock and Singh~\cite{Pollock} showing that the total entropy of gravity+matter does not decrease through physical processes.
It would be interesting to find the cosmological horizon of anisotropic universes whose geometry are not isotropic and investigate the possibility to generalize the second law of thermodynamics to such spacetimes.
At the present universe, the cosmological horizon is deeply related to dark energy which dominates the later accelerating expansion of the universe.
It is well known that the portion of the dark energy is increased during the matter dominated regime and is negligible at the radiation dominated regime.
This is why one usually neglect the dark energy contribution during inflationary and pre-inflationary regimes of the universe.
However, in the case of an holographic dark energy, one cannot neglect the dark energy portion at the pre-inflationary regime.
If we observe the universe backward in time during the inflation, we may see that the portion of the holographic dark energy density will be exponential increased until it is comparable to the inflaton energy density at the pre-inflationary universe as shown in Refs.~\cite{Chen,kjj2}.
Therefore, it is reasonable to introduce the holographic dark energy on the anisotropic inflation models and ask the possibility that the behavior of anisotropy in Kasner models is modified by its presence.
In fact, this gives a possibility to cure the pathological initial condition problem of the gravitational wave perturbation~\cite{tpu}, which happens because of the initial anisotropic singularity of Kasner models.

We are interested in the homogeneous anisotropic cosmological solutions of Einstein gravity with a positive cosmological constant, $\Lambda(=3H^2)$, in the absence of matter fields.
The spacetime solutions are known as the Bianchi universes, which are divided into nine classes.
In this paper, we study the simplest case: the Kasner spacetimes with two dimensional plane symmetry with metric,
\begin{eqnarray} \label{ds2:KdS}
ds^2=-d\tau^2+ a^2(\tau)~ (dx^2+ dy^2)+ c^2(\tau) dz^2,
\end{eqnarray}
where $a(\tau)$ and $c(\tau)$ are independent scale factors of the planes and of the direction orthogonal to the plane, respectively.
The space-time model is a good testing ground in analyzing the properties of anisotropic universes since it is simple and has various important features of the whole anisotropic universes.

There are two different solutions of distinct features.
One is given by $ds^2 := ds_{\rm r}^{2}$ with scale factors
\begin{eqnarray} \label{metric:regular}
a(\tau):= a_{\rm r}(\tau)= a_0 \sinh^{\frac13}(3H\tau) \tanh ^{-\frac{1}3}\Big(\frac{3H\tau}2\Big), \qquad
c(\tau):=c_{\rm r}(\tau) = c_0 \sinh^{\frac13}(3H\tau) \tanh ^{\frac{2}{3}}\Big(\frac{3H\tau}2\Big).
\end{eqnarray}
The spacetime has a Rindler-like event horizon at $\tau=0$ since $a_{\rm r}$ approaches to a finite value and $c_{\rm r}$ goes to zero linearly.
The spacetime is connected to the region with negative $\tau$ through the horizon where
a timelike singularity is located.
As shown in Ref.~\cite{LKKL}, the geometry is closely connected with the anti-de Sitter black brane solution discovered by Lemos~\cite{Lemos}.
The other solution, $ds^2 := ds_{\rm s}^{2}$, has an initial singularity and
their scale factors are given by
\begin{eqnarray} \label{metric:singular}
a(\tau):= a_{\rm s}(\tau)= a_0 \sinh^{\frac13}(3H\tau) \tanh ^{\frac{1}3}\Big(\frac{3H\tau}2\Big), \qquad
c(\tau):=c_{\rm s}(\tau) = c_0 \sinh^{\frac13}(3H\tau) \tanh ^{-\frac{2}{3}}\Big(\frac{3H\tau}2\Big).
\end{eqnarray}
The metric component $a_{\rm s}$ goes to zero and $c_{\rm s}$ diverges as $\tau \to 0$.
As a result, there appears a spacelike singularity, which is present in most Bianchi spacetimes other than $ds^2_{\rm r}$ style.
The solutions behave initially as Kasner spacetimes in the absence of cosmological constant followed by a de Sitter phase, as discussed in \cite{tpu,gkp}.
In this paper, we set $a_0=1=c_0$, which can always be done by an appropriate rescaling of the spatial coordinates $x^i$ ($i=1,2,3$).

Now we consider the null geodesics on this spacetime.
Consider a flash of light with conserved comoving momentum $p_i:= |\vec p|(\sin \theta \cos\phi,\sin\theta \sin \phi, \cos \theta )$ which  departed from the origin at time $\tau$, where $\theta$ denotes the angle of the comoving momentum with respect to the $z$ direction.
At time $\tau_0>\tau$, the light will arrive at the point,
\begin{eqnarray} \label{geodesic:plane}
r^i = (x,y,z);
\quad x(\tau,\tau_0) = \sin \theta \cos\phi ~\,\mathfrak{f}(\theta,\tau_0,\tau), \quad y(\tau,\tau_0) = \sin \theta \sin\phi ~\,\mathfrak{f}(\theta,\tau_0,\tau), \quad z(\tau,\tau_0) = \cos\theta \,~ \mathfrak{g}(\theta,\tau_0,\tau)
\end{eqnarray}
where
\begin{eqnarray} \label{fg}
\mathfrak{f}(\theta,\tau_0,\tau) :=\int_\tau^{\tau_0}d\tau' \frac{c(\tau') }{a(\tau')\sqrt{ \sin^2\theta \,c^2(\tau')+\cos^2\theta \,a^2(\tau')}}
,\quad
\mathfrak{g}(\theta,\tau_0,\tau):=\int _\tau^{\tau_0}d\tau'  \frac{a(\tau')}{c(\tau')\sqrt{  \sin^2\theta \,c^2(\tau')+\cos^2\theta \,a^2(\tau') }}.
\end{eqnarray}
The angle $\theta$ does not directly related to the physical zenith angle of the points $(x,y,z)$ but is related to the conserved momenta.

Note that, for large $\tau$, the two scale factors will eventually evolve exponentially and become the same as that of the de Sitter spacetime.
In this sense, the existence of the cosmological event horizon is guaranteed.
As $\tau_0\to \infty$, the light will arrive at the spacial coordinate
$$
r^i_h(\tau,\theta,\phi)=\lim_{\tau_0\to \infty} r^i(\tau,\tau_0).
$$
Conversely, a flash of light with comoving spatial momentum $-p_i$ which departed from the event $(\tau,x_h,y_h,z_h)$ will arrive at the origin as $\tau_0 \to \infty$.
This clearly implies that the surface,
\begin{eqnarray} \label{hor}
\mathfrak{H}(\tau)\equiv \{   (\tau,x_h,y_h,z_h), ~0\leq\theta\leq \pi, ~0\leq \phi <2\pi,~ \tau> 0\},
\end{eqnarray}
forms the cosmological event horizon of the universe at the comoving time $\tau$.
If $a=c$, the formula reproduce the distance to the cosmological event horizon in the Robertson-Walker spacetime.

Apparently, the horizon surface satisfies
$$
\frac{x^2+y^2}{\mathfrak{f}^2(\theta,\infty,\tau)} + \frac{z^2}{\mathfrak{g}^2(\theta,\infty,\tau)} =1,
$$
with $\mathfrak{f}$ and $\mathfrak{g}$ dependent on time and angle $\theta$. If the two functions $\mathfrak{f}$ and $\mathfrak{g}$ are independent of the angle, the surface forms an ellipse. This case happens only asymptotically as $\tau\to \infty$.
Explicitly, the distance to the cosmological event horizon is dependent on the polar angle $\theta$ and is given by
\begin{eqnarray} \label{Rh}
R_h(\tau,\theta) = \sqrt{R_\parallel^2+R_\perp^2},
\end{eqnarray}
where the distance along the parallel and orthogonal directions to the $x$-$y$ plane are given by
$$
R_\parallel(\tau,\theta)=a(\tau) \sqrt{x_h^2(\tau)+y_h^2(\tau)}= \sin\theta\, a(\tau) \mathfrak{f}(\theta,\infty,\tau),
\qquad R_\perp(\tau,\theta)=\cos\theta \, c(\tau) \mathfrak{g}(\theta,\infty,\tau).
$$

In this paper, we explicitly calculate the distances to cosmological event horizons for the two metrics~(\ref{metric:regular}) and (\ref{metric:singular}).
Let us first study the initially regular spacetime, $ds_{\rm r}^2$, case.
Considering the perturbation around the Kasner spacetime with positive cosmological constant, the gravitational wave modes rapidly grow and have instability at the initial Kasner regime, generically.
Only for the present solution, the instability can be suppressed to give a well defined initial condition for the gravitational perturbations~\cite{kM2}.
\begin{figure}[h]
\begin{center}
\includegraphics[width=.7\linewidth]{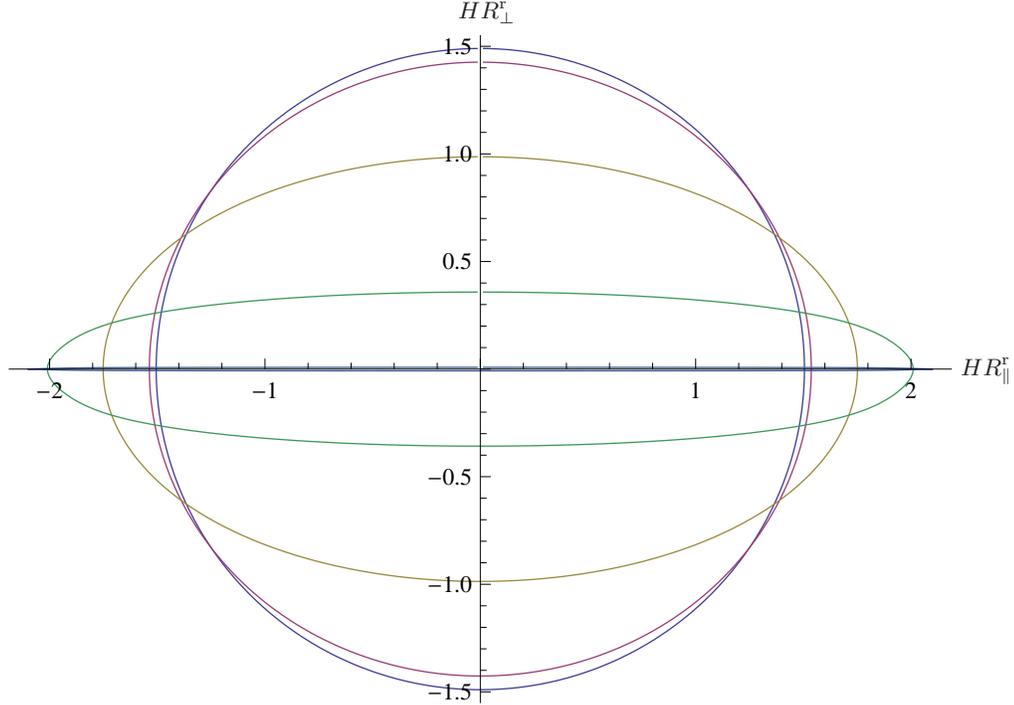}
\put (-190,260) { $HR_\perp^{\rm r}$   }
\put (0,125) { $HR_\parallel^{\rm r}$   }
\end{center}
\caption{Shape of the event horizon $ (H R_\parallel^{\rm r}, H R_\perp^{\rm r})$.
Here, $3H\tau/2=2.5,1.5,.5,.1,$ and $.001$, respectively from small eccentricity. We choose $H=2/3$.} \label{fig:eh m=-1}
\end{figure}
Eq.~(\ref{fg}) is integrable to give a closed form and the spatial distance
to the event horizon for a given angle $\theta$ becomes
\begin{eqnarray}
R_h^{\rm r}(\tau,\theta) &=& \frac{1}{H}\left( \sin^2\theta \left[f_{\rm r}(\frac{\cosh(\frac{3H\tau}{2})}{\sin\theta})\right]^2 + \cos^2\theta
\left[g_{\rm r}(\sin\theta,\cosh\ta)\right]^2\right)^{1/2},
\end{eqnarray}
where
\begin{eqnarray}
f_{\rm r}(x) &=&\frac{2}{3}x^{2/3}\int_{x}^\infty dx'\frac{1}{{x'}^{2/3}
    \sqrt{{x'}^2-1}} \nn \\
&=&\frac23 x^{2/3} \left[\frac{\sqrt{\pi}\,\Gamma(1/3)}{2\Gamma(5/6)}- \sqrt{x^2-1} ~_2F_1(1/2,5/6,3/2,
   1-x^2)\right]
\end{eqnarray}
is a monotonically decreasing function of $x~(\geq 1)$ with $f_{\rm r}(1)=
\frac{\sqrt{\pi}\Gamma(1/3)}{3\Gamma(5/6)}$ and $\displaystyle \lim_{x\to\infty} f_{\rm r}(x)=1$.
The function,
\begin{eqnarray}
g_{\rm r}(s,y) &=&   \frac{\sqrt{y^{2}-1}}{3\,y^{\frac13}}
    \int^{\infty}_{y^2-1}dy'\frac{(y'+1)^{1/6}}{y'
        \sqrt{y'+1-s^2}}
\nn \\
&=&   \frac{\sqrt{y^{2}-1}}{3\,y^{\frac13}}\sum_{n=0}^\infty \frac{\Gamma(n-1/2)}{\Gamma(1/2)\Gamma(n+1)}s^{2n} \times
    B_{y^{-2}}(1/3+n,0)
      \nn
\end{eqnarray}
monotonically increases from zero to one as $y~(\geq 1)$ increases.

For the limit $\tau\to \infty$, the distance to the horizon approaches to that of the de Sitter space as $R_h^{\rm r} \to 1/H$, and the cosmological horizon forms a spherical surface.
On the other hand, around the time $\tau \to 0$, the $z$-direction distance to the horizon goes to zero as
$$
R_\perp^{\rm r}
    \to \tau \log \frac{2}{3H\tau}.
$$
Note that this distance is independent of the angle $\theta$.
As $\tau \to 0$, the distance to the horizon parallel to the plane approaches to a finite number,
$$
R_\parallel^{\rm r}
    \to  \frac{\sqrt{\pi}|\sin^{1/3}\theta|}{3H}\frac{\Gamma(1/3)}{\Gamma(5/6)}.
$$

Therefore, the horizon take the form of a pancake surface of infinitesimal thickness as shown in the case of $3H\tau/2=0.001$ in Fig.~\ref{fig:eh m=-1}.
The area of the initial event horizon,
$$
A = 2\pi (R_{\parallel}^{\rm r})^2=\frac{2\pi^2}{9H^2} \frac{\Gamma^2(1/3)}{\Gamma^2(5/6)}\simeq \frac{12.3535}{H^2},
$$
is slightly smaller than that of the final de Sitter horizon $4\pi/H^2\simeq 12.5664/H^2$.
The horizon area are increased through the evolution.

We next consider the spacetimes with initial spacelike singularity.
Even though, the metric $ds_{\rm s}^2$ is a specific example of general Kasner spacetimes, it still has most important features of the initially singular anisotropic spacetimes.
With metric~(\ref{metric:singular}), Eq.~(\ref{fg}) is integrable to give a closed form and the spatial distance
to the event horizon for a given angle $\theta$ becomes
\begin{eqnarray}
R_h^{\rm s}(\tau,\theta) &=& \frac{1}{H}\left( \sin^2\theta \left[f_{\rm s}(\frac{\sinh(\frac{3H\tau}{2})}{\sin\theta})\right]^2 + \cos^2\theta
\left[g_{\rm s}(\cos\theta,\sinh^2\ta)\right]^2\right)^{1/2},
\end{eqnarray}
where
\begin{eqnarray}
f_{\rm s}(x) =\frac{2 x^{2/3}}{3}
   \int^{\infty}_{x} \frac{dx'}{{x'}^{2/3}\sqrt{1+{x'}^2}}=~_2F_1(\frac13,\frac12;\frac43;-\frac1{x^2}),
\end{eqnarray}
monotonically increases from zero to one as $x$ increases from zero to infinity.
The function
\begin{eqnarray} \label{g_1}
g_{\rm s}(c,y) &=& \frac{(y+1)^{1/2}}{3y^{1/6}}\int_{y}^\infty dy'
   \frac{{y'}^{1/6}}{(y'+1)^{3/2}}\left(1-\frac{c^2}{y'+1}\right)^{-1/2} \nn \\
&=& \frac{(y+1)^{1/2}}{3y^{1/6}} \sum_{n=0}^\infty \frac{\Gamma(n+1/2)}{\Gamma(1/2)\Gamma(n+1)} c^{2n} (-)^{5/3+n}
    B_{-y^{-1}}(1/3+n,-1/2-n),
\end{eqnarray}
monotonically decreases from infinity to one as $y~(>0)$ increases.
As a result, the distance to the horizon asymptotically approaches to that of the de Sitter space as $R_h^{\rm s} \to 1/H$.

\begin{figure}[h]
\begin{center}
\includegraphics[width=.4\linewidth]{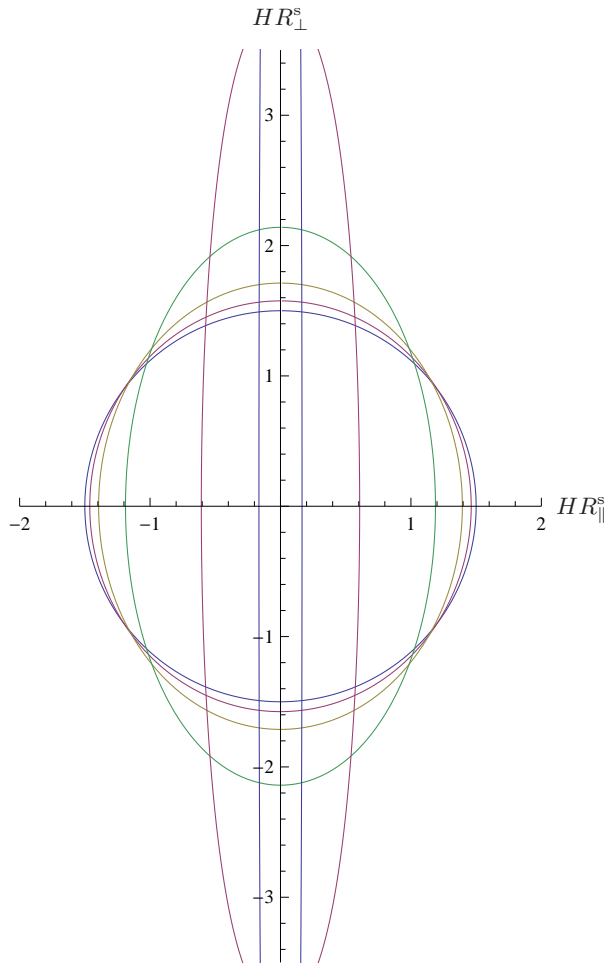}
\put (-115,355) { $HR_\perp^{\rm s}$   }
\put (0,170) { $HR_\parallel^{\rm s}$   }
\end{center}
\caption{Shape of the event horizon $ (H R_\parallel^{\rm s}, H R_\perp^{\rm s})$.
Here, $3H\tau/2=1.5,1,.5,.1$ and $.01$, respectively from small eccentricity.}
\label{fig:eh m=1}
\end{figure}
On the other hand, around the initial singularity, the $z$-direction distance to the horizon diverges as
$$
 R_\perp^s \to \frac{\Gamma(4/3)\Gamma(1/6)}{\sqrt{\pi} H}  \frac{\cos\theta \, _2F_1(1/3,1/2,3/2,\cos^2\theta)}{3\sinh^{1/3}\ta}.
$$
The distance to the horizon parallel to the plane approaches to zero as
$$
R_\parallel^s \to \frac{\Gamma(4/3)\Gamma(1/6)}{\sqrt{\pi} H}
\sin^{1/3}\theta \sinh^{2/3}\ta.
$$
Therefore, the initial horizon take the form of a needle surface of infinitesimal thickness and infinite length as shown in Fig.~\ref{fig:eh m=1}.
The area of the initial cosmological horizon goes to zero as
$$
A\sim 2\pi R_{\parallel}^s R_\perp^s \propto \frac{\sinh^{1/3}\ta}{H^2} .
$$
The horizon area is increased through the evolution.
We plot the event horizons at several cosmological times in Fig.~\ref{fig:eh m=1}.
The whole form of the event horizon can be obtained by rotating the graph around the $z$-axis (the horizontal axis).
For $H\tau \gg 1$ the event horizon becomes  a spherical surface.
The eccentricity is very high initially even though its shape is not an exact ellipsoid.

Let us summarize the results in this paper.
At the future infinity, the cosmological event horizon has spherical shape and is located at $1/H$.
For times other than the future infinity, it is non-spherical and has non-zero eccentricity, which is maximized at the initial time.
In the case of the initially singular Kasner spacetimes, it was shown that the initial cosmological horizon takes the form of a needle surface with infinite length and infinitesimal thickness.
The initial surface area goes to zero since the thickness decreases faster than the increase of the length if we observe the universe backward in time around the initial singularity.
In the case of the initially regular Kasner spacetime, the initial cosmological horizon takes the form of a pancake surface on the symmetric plane with infinitesimal thickness.
For both cases, the surface areas of the cosmological event horizons increase with time.
Therefore, the $2^{\rm nd}$ law of thermodynamics holds in the anisotropic universe allowing the entropy interpretation of the horizon area.

Calculating surface gravity on the horizon is a challengeable task since the Killing vector generating the horizon is not known.
The vector following the horizon for a given value of $\theta$ and $\phi$, are given by
$$
v^a = \frac{d x_h^a}{d\tau} = (1,-\frac{c}{a}\frac{\sin\theta \cos\phi }{\sqrt{\cos^2\theta\,a^2+\sin^2\theta\,c^2}}, -\frac{c \sin\theta\sin\phi}{a\sqrt{\cos^2\theta\,a^2+\sin^2\theta\,c^2}},  -
    \frac{a \cos\theta}{c \sqrt{\cos^2\theta\,a^2+\sin^2\theta\,c^2}}).
$$
One can easily show that this is null $v^a v_a=0$ but not a Killing vector. The horizon generating Killing vector $\chi_a$ will be proportional to $v_a$ at the horizon. We defer this subject of finding the vector to later work.
Rather, we roughly estimate the surface gravity near $\tau\sim 0$ by using analogy to the de Sitter spacetime.
Near the cosmological horizon located at $x=H^{-1} e^{-H\tau}$ with $y=0=z$, the metric of the de Sitter spacetime becomes
$$
ds^2 \simeq -2 a(\tau) d\tau d\xi+ a^2(\tau)( d\xi^2 +dy^2+dz^2),
$$
where $\xi \equiv x- H^{-1}a^{-1}$ and $a(\tau) = e^{H\tau}$. The surface $\xi=0$ define the cosmological horizon. The surface gravity in this spacetime is given by the derivative of the logarithm of scale factor, $a^{-1}(d a/d\tau)$.
In this calculation, the direction $y$ and $z$ are irrelevant.

In the anisotropic Kasner spacetimes, we may proceed in a similar manner to estimate the surface gravity of the cosmological horizons at early times.
For the case of the initially singular metric $ds_{\rm s}^2$, most part of the horizon is almost parallel to the $z$-axis initially since the horizon has needle-like shape.
Therefore, we may assume rough translational symmetry along $z$.
We estimate the surface gravity at the event $(\tau\ll H^{-1}, x_h, y_h=0, z_h\sim 0)$ on the horizon.
Since the geometry is almost needle-like, we may simply calculate the surface gravity at $\theta = \pi/2$.
Similar calculation as above leads to $\kappa_{\rm s} \sim H_a = \dot a/a \propto 1/\tau $ at the early times.
For the case of the initially regular metric $ds_{\rm r}^2$, the horizon is almost pancake shape located on the $x$-$y$ plane initially.
We may estimate the surface gravity at the point $\theta =0$.
Similar calculation leads to $\kappa_{\rm r} \sim H_c = \dot c/c \propto 1/\tau $.

The surface gravity does the role of a Hawking temperature for the geometry.
Usually, a kind of thermal energy can be related to the horizon and at the present case  one usually relate it to the holographic dark energy~\cite{Li,jkj} which explains the cosmic expansion of our universe.
It was also pointed out that the second law of thermodynamics holds,
$
\frac{dA_{EH}}{dt} \geq 0,
$
for the case of a Robertson-Walker cosmology with holographic dark energy.
In addition, the second law can be used to determine when the dark energy dominated era begins in the universe~\cite{hjj}.

Let us assume that there present small amount of holographic dark energy and matter in the later (but before inflation) evolution on the background of Kasner spacetime.
Then, let us see how does the energy densities behave as we go backward in time.
The holographic dark energy density at earlier times behaves as
$$
\rho_H \sim \frac{T_H S}{V_{\rm in}}\sim   \frac{\kappa A}{R_\parallel^{~2}(\tau,\pi/2) R_\perp(\tau,0)},
$$
where $T_H$, $S$, and $V_{\rm in}$ denote the Hawking temperature, entropy, and volume of the universe inside the horizon, respectively.
Therefore, the holographic dark energy density in the initially regular or singular spacetimes become
$$
\rho_{H,{\rm r}} \sim \frac{H^2}{(H\tau)^3}\left(\log \frac{2}{3H\tau}\right)^{-2}, \qquad
\rho_{H,{\rm s}} \sim \frac{H^2}{(H\tau)^{7/3}}.
$$
The energy density of matter field behaves as
$ \rho_{m,{\rm r}} \sim \rho_0/(a^2(\tau) c(\tau)) \sim \rho_0/(H\tau)$ and $\rho_{m,{\rm s}} \sim \rho_0/(H\tau)^{2/3}$, respectively.
As time goes back, the holographic dark energy density increases much faster than that of the matter.
Therefore, the holographic dark energy eventually dominates the universe at some early times.
If one consider the reaction of the holographic dark energy to background geometry, which is not taken care of in this paper, the evolution of the Kasner universe should be modified.

\section*{Acknowledgement}
HCK was supported in part by the Korea Science and Engineering Foundation
(KOSEF) grant funded by the Korea government (MEST) (No.2010-0011308) and the APCTP Topical Research Program (2011-T-01).


\end{document}